\documentclass[prd,twocolumn,nofootinbib]{revtex4}
\usepackage{graphicx, epsfig}
\usepackage{color}
\usepackage{mathrsfs}
\usepackage{bm}
\usepackage{amsmath,amssymb,empheq}% for \eqref
\usepackage{mathrsfs}
\usepackage[caption=false]{subfig}
\usepackage{hyperref}
\usepackage[normalem]{ulem}
\usepackage{mathtools}
\usepackage[x11names]{xcolor}

\definecolor{fashionfuchsia}{rgb}{0.96, 0.0, 0.63}
\colorlet{no_so_fashion_purple}{blue!50!red}

\newcommand{\be}{\begin{equation}}
\newcommand{\ee}{\end{equation}}
\newcommand{\ba}{\begin{eqnarray}}
\newcommand{\ea}{\end{eqnarray}}

\newcommand{\nn}{\nonumber}

\begin{document}
\title{Electric flux tube solutions in SU(3) gauge theory}
\author{Jude Pereira\footnote{jperei10@asu.edu}, Tanmay Vachaspati\footnote{tvachasp@asu.edu}}
\affiliation{
Physics Department, Arizona State University, Tempe,  Arizona 85287, USA.
}

\begin{abstract}
We consider electric flux tube solutions in SU(3) gauge theory with scalar fields in the
fundamental representation. Such solutions can possibly be constructed in two classes, 
corresponding to the two maximally commuting generators $\lambda_3$ and $\lambda_8$ 
of SU(3). We successfully construct the 3-type solution with two scalar fields and show that 
it is immune to decay through Schwinger pair production of gauge bosons but we have not been able to construct an 8-type solution.
\end{abstract}

\maketitle

\section{Introduction}
\label{intro}

Magnetic type solutions in non-Abelian gauge theories, including magnetic monopoles and cosmic strings,
have been of interest for many decades. Electric-type solutions, however, have only recently been 
discovered and their existence and properties remain largely unexplored. The key to the new electric
string solutions is the discovery by Brown and Weisberger (BW)~\cite{BROWN1979285} that there are an 
infinite number of gauge inequivalent non-Abelian gauge fields, in addition to embedded 
Maxwellian gauge fields, all of which lead to identical electric fields. Electric fields resulting from
Maxwellian gauge fields are vacuum solutions of the equations of motion but such electric fields
are unstable to rapid Schwinger pair production of gauge particles and will eventually 
dissipate~\cite{Casher1979,Nayak2005,Cardona:2021ovn}. 
On the other hand, electric fields generated from the BW gauge fields need volume-filling sources
but are classically stable for a wide range of parameters~\cite{Pereira:2024sic, Pereira:2022lbl}
and also quantumly stable (``unexciting'') to Schwinger 
pair production of gauge particles~\cite{Vachaspati:2022ktr}.

Here we are interested in BW gauge fields that result in electric-type solutions.
Since such solutions require non-trivial sources, it is necessary to include ``matter'' fields 
that are charged under the non-Abelian symmetry.
In Ref.~\cite{Vachaspati:2022gco,Vachaspati:2023tpt} electric-type solutions were constructed in 
an SU(2) non-Abelian gauge theory with the matter field chosen to be a scalar field transforming 
in the fundamental representation of SU(2). The ``electric string'' solution, equivalently
an ``electric flux tube'', was derived using some guesswork and it seemed somewhat of a miracle 
that all the equations of motion could be solved for this highly non-linear system. The question
we address in this paper is if the same miracle can be extended to other gauge groups, and
SU(3) in particular because of its relevance to quantum chromodynamics (QCD).

At first sight, it appears that the SU(2) solution can simply be embedded into SU(3). However,
this turns out not to be the case for the reason that the SU(2) Pauli spin matrices, $\sigma^a$,
($a=1,2,3$), satisfy
\be
\{ \sigma^a,\sigma^b \} = 2\delta^{ab}
\ee
whereas the SU(3) Gell-Mann matrices, $\lambda^a$, ($a=1,\ldots,8$) satisfy,
\be
\{ \lambda^a,\lambda^b \} = \frac{4}{3} \delta^{ab} + 2 \, d^{abc} \lambda^c
\ee
where $d^{abc}$ are non-vanishing in general. 
Further, SU(3) has two elements in its Cartan subalgebra and there are two different
``electric fields'', one in the $\lambda^3$ direction and the other in the $\lambda^8$
direction. Thus there are two types of electric solutions that one can consider in
SU(3). We shall call these ``3-type'' and ``8-type'' solutions.

Construction of the 3-type electric field largely follows the procedures developed for
the SU(2) case~\cite{Vachaspati:2022gco,Vachaspati:2023tpt} except that we find
that a single scalar field cannot source the BW gauge fields. Instead we need to
introduce two scalar fields that transform in the fundamental representation of SU(3).
With such matter fields we can source a uniform electric field in the $\lambda^3$
direction as well as an electric string solution (see Section~\ref{seclambda3}). Construction of the 8-type electric field turns out to be more complicated since
$d^{abc} \ne 0$ and will be reserved for future work.

One motivation for investigating the BW gauge fields is that they are quantumly
stable. This is shown explicitly for the 3-type electric field in Sec.~\ref{unexciting}.

We start out in Sec.~\ref{gaugefields} by describing 3-type and 8-type BW gauge 
fields for SU(3) and their field strengths. In Section~\ref{seclambda3} we analyze 
3-type electric solutions with two scalar fields in the fundamental representation and summarize the solution in Sec.~\ref{summary}. We discuss questions for future work in the concluding section~\ref{conclusions}. We also show in Appendix~\ref{appC} that a single scalar field in the adjoint representation of SU(3) cannot provide suitable sources.

\section{Gauge fields}
\label{gaugefields}

SU(3) gauge fields, $W_\mu^a$, of the Maxwellian type that give rise to an electric field in the
$z$-direction in space and in the
$\lambda^b$ direction in group space may be written as
\be
W_\mu^a = - E t \, \partial_\mu z \, \delta^{a b}.
\ee
In contrast, a general set of BW gauge fields that result in uniform electric fields along the
$z$-direction and in the $a=3$ and $a=8$ directions in group space can be written as
\be\label{BWfields}
\begin{aligned}
  W^{1}_{\mu} &= \frac{\Omega_1}{g}\partial_{\mu}t, \quad W^{2}_{\mu} = -\frac{\Omega_1 E}{\Omega^2}\partial_{\mu}z, \quad W^{4}_{\mu} = \frac{\Omega_2}{g}\partial_{\mu}t, \\
W^{5}_{\mu} &= \frac{\Omega_2 E}{\Omega^2}\partial_{\mu}z, \quad W^{6}_{\mu} = \frac{\Omega_3}{g}\partial_{\mu}t, \quad W^{7}_{\mu} = - \frac{\Omega_3 E}{\Omega^2}\partial_{\mu}z  
\end{aligned}
\ee
Using the expression for the field strength tensor
\be\label{fieldstrength}
W^{a}_{\mu\nu} = \partial_{\mu}W^{a}_{\nu} - \partial_{\nu}W^{a}_{\mu} + g f^{abc} W^{b}_{\mu}W^{c}_{\nu}
\ee
where $f^{abc}$ are the $SU(3)$ structure constants defined using the commutator of the Gell-Mann 
matrices~\cite{Haber}
\be
[\lambda_a, \lambda_b] = 2if^{abc}\lambda_c
\ee
we obtain
\ba
W^{3}_{\mu\nu} = \frac{E}{2\Omega^2}(\Omega_2^2 + \Omega_3^2 - 2\Omega_1^2)(\partial_{\mu}t\partial_{\nu}z - \partial_{\mu}z\partial_{\nu}t) 
\label{W3munu}
\\
W^{8}_{\mu\nu} = \frac{\sqrt{3}E}{2\Omega^2}(\Omega_2^2 - \Omega_3^2)(\partial_{\mu}t\partial_{\nu}z - \partial_{\mu}z\partial_{\nu}t)
\label{W8munu}
\ea
Setting $\Omega_2^2 = \Omega_3^2$ results in a constant electric field in the $a=3$ direction (3-type) whereas setting $\Omega_1^2 = (\Omega_2^2 + \Omega_3^2)/2$ results in a constant electric field in the $a=8$ direction (8-type).

The gauge four-currents are defined as
\be\label{fourcurrent}
j^{\mu a} \equiv D_{\nu}W^{\mu\nu a} = \partial_{\nu}W^{\mu\nu a} + g f^{abc} W^{b}_{\nu}W^{\mu\nu c}
\ee
which can be computed for the above BW gauge fields as follows
\ba
j^{1}_{\mu} &=& \frac{gE^2}{2\Omega^4}\Omega_1(\Omega_2^2 + \Omega_3^2 - 2\Omega_1^2)\partial_{\mu}t \\
j^{2}_{\mu} &=& \frac{E}{2\Omega^2}\Omega_1(\Omega_2^2 + \Omega_3^2 - 2\Omega_1^2)\partial_{\mu}z \\
j^{4}_{\mu} &=& \frac{gE^2}{2\Omega^4}\Omega_2(\Omega_1^2 + \Omega_3^2 - 2\Omega_2^2)\partial_{\mu}t \\
j^{5}_{\mu} &=& -\frac{E}{2\Omega^2}\Omega_2(\Omega_1^2 + \Omega_3^2 - 2\Omega_2^2)\partial_{\mu}z \\
j^{6}_{\mu} &=& \frac{gE^2}{2\Omega^4}\Omega_3(\Omega_1^2 + \Omega_2^2 - 2\Omega_3^2)\partial_{\mu}t \\
j^{7}_{\mu} &=& \frac{E}{2\Omega^2}\Omega_3(\Omega_1^2 + \Omega_2^2 - 2\Omega_3^2)\partial_{\mu}z
\ea
with other components vanishing.

In what follows, we will investigate only 3-type electric fields and attempt to construct electric string solutions following \cite{Vachaspati:2022gco,Vachaspati:2023tpt}.

\section{Electric field in $\lambda_3$ direction}
\label{seclambda3}

\subsection{Homogeneous electric fields}
\label{homogeneousE}

Electric fields in the $\lambda_3$ direction may be obtained by setting $\Omega_2^2 = \Omega_3^2 =0$. 
We also denote $\Omega_1 = \Omega$. This results in the following gauge fields
\be
W^{1}_{\mu} = \frac{\Omega}{g}\partial_{\mu}t, \quad W^{2}_{\mu} = -\frac{E}{\Omega}\partial_{\mu}z,
\ee
with all other $W^{b}_{\mu} = 0$. The only non-vanishing field strength component is
\be
W^{3}_{\mu\nu} = -E(\partial_{\mu}t\partial_{\nu}z - \partial_{\mu}z\partial_{\nu}t)
\ee
These gauge fields can be written in the temporal gauge $W^{a}_t = 0$ by acting with a gauge transformation $U$
\be
\label{gaugetransform}
W_{\mu} \to W'_{\mu} = U W_{\mu}U^{\dagger} + \frac{2i}{g}U\partial_{\mu}U^{\dagger}
\ee
where $W_{\mu} = W^{a}_{\mu}\lambda_{a}$. Following \cite{Vachaspati:2023tpt}, it can be checked that
\be
U = e^{i \lambda_2 \pi/4} e^{i\lambda_1 \pi/4} e^{-i\lambda_1\Omega t/2}
\ee
results in
\be
\label{gaugeField}
W^{1}_{\mu} = -\frac{\epsilon}{g}\cos{\Omega t}\partial_{\mu}z, \ \ 
W^{2}_{\mu} = -\frac{\epsilon}{g}\sin{\Omega t}\partial_{\mu}z
\ee 
along with $W^{b}_{\mu} = 0$ for $b = 3,\ldots, 8$ where we have defined
\be
\epsilon = \frac{gE}{\Omega} .
\ee
The non-vanishing field strength tensor for the gauge fields in the temporal gauge can be shown to be
\ba
W^{1}_{\mu\nu} &=& \frac{\epsilon\Omega}{g}\sin(\Omega t)\, (\partial_{\mu}t\partial_{\nu}z - \partial_{\nu}t\partial_{\mu}z) \\
W^{2}_{\mu\nu} &=& -\frac{\epsilon\Omega}{g}\cos(\Omega t)\, (\partial_{\mu}t\partial_{\nu}z - \partial_{\nu}t\partial_{\mu}z) 
\ea
Note that the electric field $W^{a}_{0i}$ is time-dependent in the temporal gauge. However, the gauge-invariant energy density in the electric fields
\be
\mathcal{E}_{\text{elec}} = \frac{1}{2}(W^{a}_{0i})^2 = \frac{\epsilon^2\Omega^2}{2g^2} = \frac{1}{2}E^2
\ee
is time-independent.

Using \eqref{fourcurrent}, the four-currents for the gauge fields in the temporal gauge can be computed as follows 
\ba
j^{1}_{\mu} &=& - \frac{\epsilon\Omega^2}{g}\cos(\Omega t)\partial_{\mu}z 
\label{gaugecurrent31}
\\
j^{2}_{\mu} &=& - \frac{\epsilon\Omega^2}{g}\sin(\Omega t)\partial_{\mu}z 
\label{gaugecurrent32}
\\
j^{3}_{\mu} &=& - \frac{\epsilon^2\Omega}{g}\partial_{\mu}t
\label{gaugecurrent33}
\ea
while other components vanish.

\subsection{Scalar fields}
\label{type3scalar}
Next, we would like to obtain scalar fields that can consistently source the gauge fields
in \eqref{gaugeField}. 
We consider the following Lagrangian for the $SU(3)$ gauge fields coupled to 
two\footnote{We will see later that a single scalar field cannot source the gauge fields.} 
scalar fields in the fundamental representation \cite{Rubakov:2002fi}
\be
L = -\frac{1}{4}W^{a}_{\mu\nu}W^{\mu\nu a} + |D_{\mu}\Phi_1|^2 + |D_{\mu}\Phi_2|^2 - V(\Phi_1,\Phi_2)
\label{lag}
\ee
where
\be\label{covDev}
D_{\mu}\Phi_i = \partial_{\mu}\Phi_i - i\frac{g}{2}W_{\mu}^{a}\lambda^{a}\Phi_{i} ,
\ee
for $i=1,2$ and
\begin{equation}
\begin{aligned}
     V(\Phi_1,\Phi_2) &= m_1^2 |\Phi_1|^2 + m_2^2 |\Phi_2|^2 + \kappa_1 |\Phi_1|^4 \\&+ \kappa_2 |\Phi_2|^4 + \kappa_{12} \big[(\Phi_1^{\dagger}\Phi_2)^2 + (\Phi_2^{\dagger}\Phi_1)^2\big]
\end{aligned}
\label{potentialV}
\end{equation}
The equation of motion for the scalar field $\Phi_i$ is
\be
D_{\mu}D^{\mu}\Phi_i + \delta_i V(\Phi_1,\Phi_2) = 0, \quad i=1,2
\ee
where $\delta_i$ denotes differentiation with respect to $\Phi_i$.
We evaluate
\ba\label{covDcovDPhi}
  D_{\mu}D^{\mu}\Phi_i &=& \partial_t^2\Phi_i + \frac{g^2}{4}W^a_zW^b_z\lambda^a\lambda^b\Phi_i \nn \\ 
  && \hskip -1.5 cm
  = \partial_t^2\Phi_i + \frac{g^2}{6}\big(W^a_z\big)^2\Phi_i 
  + \frac{g^2}{4}W^a_zW^b_zd^{abc}\lambda^c\Phi_i
\ea
where we have used the temporal gauge $W^a_t = 0$ and $\Phi_i = \Phi_i(t)$ to cancel out 
and simplify terms in the first line, and used the $SU(3)$ identity
\be
\lambda^a\lambda^b = \frac{2}{3}\delta^{ab} + d^{abc}\lambda^c + i f^{abc}\lambda^c
\label{lambdalambda}
\ee 
along with the total antisymmetry of the structure constants,  $f^{abc}$, to get the second line.
We also have
\be
\delta_iV(\Phi_1,\Phi_2) = m_i^2\Phi_i + 2 \kappa_i |\Phi_i|^2\Phi_i + 2\kappa_{ij}(\Phi_i^{\dagger}\Phi_j)\Phi_j
\ee
where $j\neq i$.

Let us denote $M \equiv g^2W^a_zW^b_zd^{abc}\lambda^c/4$ where repeated indices $a,b,c = 1,\ldots,8$ are summed over so that $M$ is a $3\times3$ matrix. We can use the eigenvectors of $M$ to construct an orthonormal basis for the set of scalar fields. 

For the gauge fields in \eqref{gaugeField}, we have 
\be
M = \frac{g^2}{4}\Big(\big(W^{1}_{z}\big)^2d^{118}\lambda^8 + \big(W^{2}_{z}\big)^2d^{228}\lambda^8\Big) = \frac{\epsilon^2}{2\sqrt{3}}\lambda^8
\ee
which has the orthonormal eigenvectors
\be
\psi_1 = \begin{pmatrix}
    1\\ 0\\ 0
\end{pmatrix},
\quad
\psi_2 = \begin{pmatrix}
    0\\ 1\\ 0
\end{pmatrix},
\quad
\psi_3 = \begin{pmatrix}
    0\\ 0\\ 1
\end{pmatrix}
\ee
with eigenvalues $\epsilon^2/12,\ \epsilon^2/12$ and $-\epsilon^2/6$ respectively. As we will show below, it turns out that a solution can be obtained simply by writing the two scalar fields as a linear combination of $\psi_1$ and $\psi_2$.

We write down an ansatz for the two scalar fields using the Hopf parametrization as follows
\ba
\Phi_1 &=& \eta_1\big(\cos{\alpha_1}\, e^{i\beta_1}\psi_1 + \sin{\alpha_1}\, e^{i\gamma_1}\psi_2\big) \nn\\
&=&   \eta_1\begin{pmatrix}
    \cos{\alpha_1}\, e^{i\beta_1}\\ \sin{\alpha_1}\, e^{i\gamma_1}\\ 0
\end{pmatrix},
\\
\Phi_2 &=& \eta_2\big(\cos{\alpha_2}\, e^{i\beta_2}\psi_1 + \sin{\alpha_2}\, e^{i\gamma_2}\psi_2\big) \nn\\
&=& \eta_2\begin{pmatrix}
    \cos{\alpha_2}\, e^{i\beta_2}\\ \sin{\alpha_2}\, e^{i\gamma_2}\\ 0
\end{pmatrix}
\ea
where $\alpha_1, \alpha_2, \beta_1, \beta_2, \gamma_1, \gamma_2$ are functions of $t$ while 
$\eta_1, \eta_2$ are constants.

We will further require the two scalar fields to be orthogonal. This ensures that their corresponding interaction term in the potential \eqref{potentialV} vanishes. Demanding $\Phi_1^{\dagger}\Phi_2 = 0$ gives
\be
\cos{\alpha_1}\cos{\alpha_2}e^{i(\beta_2 - \beta_1)} + \sin{\alpha_1}\sin{\alpha_2}e^{i(\gamma_2 - \gamma_1)} = 0
\ee
which can be further simplified as follows
\be
\cot{\alpha_1}\cot{\alpha_2}e^{i(\beta_2-\beta_1-\gamma_2+\gamma_1)} = -1
\ee
Since the RHS is real-valued, we require that
\be
\beta_2-\beta_1-\gamma_2+\gamma_1 = n \pi , \ n \in \mathbb{Z}
\label{beta2eq}
\ee
This results in
\be
\cot{\alpha_1}\cot{\alpha_2} = (-1)^{n+1}
\ee
which can be solved by 
\be
\alpha_1 + (-1)^{n+1}\alpha_2 = (2m - 1)\pi/2 , \ m\in\mathbb{Z}
\label{alpha1eq}
\ee
For the sake of simplicity, we set $m=0$ and $n=0$ to obtain
\ba
\beta_2 - \beta_1 = \gamma_2 - \gamma_1 \equiv \theta \\
\alpha_2 = \alpha_1 + \pi/2
\ea
where it is sufficient to consider $\theta$ to be a constant.

Thus, the final set of orthogonal scalar fields $\Phi_1$ and $\Phi_2$
\be \label{phiansatz}
\Phi_1 = \eta_1\begin{pmatrix}
    \cos{\alpha}\, e^{i\beta}\\ \sin{\alpha}\, e^{i\gamma}\\ 0
\end{pmatrix},
\quad
\Phi_2 = \eta_2e^{i\theta}\begin{pmatrix}
    -\sin{\alpha}\, e^{i\beta}\\ \cos{\alpha}\, e^{i\gamma}\\ 0
\end{pmatrix}
\ee
where we have denoted $\alpha_1 \equiv \alpha,\ \beta_1 \equiv \beta$ and $\gamma_1 \equiv \gamma$.\\

The four-current for the two scalar fields is
\be
J^{a}_{\mu} = \frac{i g}{2} \sum_{i=1,2}\big[\Phi_i^{\dagger}\lambda^{a}D_{\mu}\Phi_i - \text{h.c.}\big]
\ee
Using the definition of the gauge covariant derivative \eqref{covDev} and the $SU(3)$ identity \eqref{lambdalambda}, we can expand the terms in the four-current as follows
\ba
\Phi_i^{\dagger}\lambda^aD_{\mu}\Phi_i &=& 
\Phi_i^{\dagger}\lambda^a\partial_{\mu}\Phi_i 
- \frac{i g}{2}W^b_{\mu}\Phi^{\dagger}_i\lambda^a\lambda^b\Phi_i \nn \\
 && \hskip - 1.5 cm
 = \Phi^{\dagger}_i\lambda^a\partial_{\mu}\Phi_i - \frac{ig}{3}W^a_{\mu}|\Phi_i|^2 
 +\frac{g}{2}f^{abc}W^b_{\mu}\big(\Phi_i^{\dagger}\lambda^c\Phi_i\big) 
 \nn \\ && \hskip 1 cm
 - \frac{ig}{2}d^{abc}W^{b}_{\mu}\big(\Phi_i^{\dagger}\lambda^c\Phi_i\big)
\ea
Hence, the final expression for the four-current becomes
\ba
\label{scalarcurrent}
    J^{a}_{\mu} &=& 
   \sum_{i=1,2}\Big\{\frac{i g}{2} \big[\Phi_i^{\dagger}\lambda^{a}\partial_{\mu}\Phi_i - \text{h.c.}\big]
   \nn \\ 
   &&
   +\frac{g^2}{3}W^{a}_{\mu}|\Phi_i|^2 + \frac{g^2}{2}d^{abc}W^{b}_{\mu}\big(\Phi_i^{\dagger}\lambda^c\Phi_i\big)\Big\}
\ea

The non-vanishing components of the four-current for the set of scalar fields 
in \eqref{phiansatz} can now be computed. 
Setting $\mu = z$ in \eqref{scalarcurrent} we obtain
\be\label{scalarcurrentz}
    J^{a}_{z} = \sum_{i=1,2}\Big\{\frac{g^2}{3}W^{a}_z|\Phi_i|^2 + \frac{g^2}{2}d^{abc}W^{b}_z\big(\Phi_i^{\dagger}\lambda^c\Phi_i\big)\Big\}
\ee
Substituting \eqref{phiansatz} in \eqref{scalarcurrentz}, we obtain the $a=8$ component
\be\label{scalarcurrent8z}
    J^{8}_{z} = -\frac{g \epsilon}{2\sqrt{3}}(\eta_1^2 - \eta_2^2) \cos{(\Omega t + \beta-\gamma)}\sin{2\alpha}
\ee
This must vanish to match the corresponding gauge field four-current components that are
given in \eqref{gaugecurrent31}-\eqref{gaugecurrent33}.
Then we can either set $\eta_1 = \eta_2$ or set $\sin (2\alpha)=0$. The latter choice turns out to be inconsistent with the condition $j^8_t=0$ and so we are forced to set $\eta_1 = \eta_2 \equiv \eta$. Note that both $\eta_1$ and $\eta_2$ have to be non-vanishing, which means that our construction would fail if we only searched for a solution with a single scalar field.

The non-trivial z-component of the scalar field four-current can be written as
\be
J^{a}_{z} = g^2 \eta^2 W^a_{z}
\ee
Comparing with \eqref{gaugecurrent31}-\eqref{gaugecurrent32}, we obtain
\be
\eta^2 = \frac{\Omega^2}{g^2} \implies \eta = \frac{\Omega}{g}
\ee

Setting $\mu = t$, we get
\be\label{scalarcurrentt}
J^{a}_{t} = \frac{i g}{2}\sum_{i=1,2} \big[\Phi_i^{\dagger}\lambda^{a}\dot{\Phi}_i - \dot{\Phi}_i^{\dagger}\lambda^{a}\Phi_i\big]
\ee
where overdots denote time derivatives.
Substituting \eqref{phiansatz} in \eqref{scalarcurrentt} and setting $\eta_1 = \eta_2 = \eta$, the non-trivial t-component of the scalar field four-current is
\be\begin{aligned}
    J^1_t &= 2g \eta^2 \sin{(\beta-\gamma)}\dot{\alpha}\\
    J^2_t &= 2g \eta^2 \cos{(\beta-\gamma)}\dot{\alpha}\\
    J^3_t &= -g\eta^2(\dot{\beta} - \dot{\gamma})\\
    J^8_t &= -\frac{g\eta^2}{\sqrt{3}}(\dot{\beta}+\dot{\gamma})
\end{aligned}\ee
Comparing with  \eqref{gaugecurrent33}, we see that $J^1_t$ and $J^2_t$ can be made to vanish by setting 
$\dot{\alpha} = 0$, whereas $J^8_t$ can be shown to vanish by demanding 
$\dot{\gamma} = -\dot{\beta}$. Matching the remaining non-trivial component $J^3_t$ yields
\be
2 g^2\eta^2 \dot{\beta} = \epsilon^2\Omega \implies \dot{\beta} = \frac{\epsilon^2}{2\Omega}
\ee
where we have substituted $\eta^2 = \Omega^2/g^2$. Thus, the final set of orthogonal scalar field solutions can be written as
\be\label{finalphi}
\Phi_1 = \eta\begin{pmatrix}
    z_1 e^{i\omega t}\\ z_2 e^{-i\omega t}\\ 0
\end{pmatrix},
\quad
\Phi_2 = \eta\begin{pmatrix}
    w_1 e^{i\omega t}\\ w_2 e^{-i\omega t}\\ 0
\end{pmatrix}
\ee
where $z_1, z_2, w_1, w_2 \in \mathbb{C}$ are constants satisfying the following constraints
\be
\label{constraints}
\begin{aligned}
    |z_1|^2 + |z_2|^2 &= 1 \\ 
    |w_1|^2 + |w_2|^2 & = 1 \\
    z_1 w_1^{*} + z_2 w_2^{*} &= 0
\end{aligned}
\ee
and we have defined
\be\label{omegadef}
\omega = \frac{\epsilon^2}{2\Omega}
\ee
Note that even though we made the simplifying assumption $m = n = 0$
(see \eqref{beta2eq} and \eqref{alpha1eq}), the final solution, as written above, 
is valid generally requiring only that the two scalar fields be orthogonal. 

Thus far we have demanded that the scalar field four-current, $J^a_\mu$, matches the gauge field 
four-current, $j^a_\mu$, and in doing so we have satisfied the gauge field equations of motion. 
Next, we need to demand that both $\Phi_1$ and $\Phi_2$ satisfy their own equations 
of motion
\be
D_{\mu}D^{\mu}\Phi_i + V'(\Phi_i) = 0
\label{Phieq}
\ee
where now
\be
V(\Phi_i) = m_i^2|\Phi_i|^2 + \kappa_i|\Phi_i|^4
\ee
as there are no interaction terms since the two scalar fields are orthogonal $\Phi_1^{\dagger} \Phi_2 = \Phi_2^{\dagger} \Phi_1 = 0$. Substituting \eqref{finalphi} in \eqref{covDcovDPhi}, we obtain
\be
  D_{\mu}D^{\mu}\Phi_i = -\big(\omega^2-\epsilon^2/4\big)\Phi_i
\ee
We also have
\be\begin{aligned}
    V'(\Phi_i) &= m_i^2\Phi_i + 2\kappa_i |\Phi_i|^2\Phi_i \\
&= (m_i^2 + 2\kappa_i\eta^2) \Phi_i
\end{aligned}
\ee
Putting everything together, \eqref{Phieq} gives
\be
-\omega^2 + \frac{\epsilon^2}{4} + m_i^2+ 2\kappa_i\eta^2 = 0
\ee
Substituting $\epsilon^2 = 2\omega\Omega$ from \eqref{omegadef} and solving the resulting quadratic equation, we obtain
\be
\omega = \frac{1}{2}\Bigg[\frac{\Omega}{2}\pm \sqrt{\frac{\Omega^2}{4}+4\Big(m_i^2 + \frac{2\kappa_i}{g^2}\Omega^2\Big)} \, \Bigg]
\ee
Since a single value of $\omega$ must satisfy the equation of motion for both scalar fields, this implies additional constraints on the model parameters. In particular, we must have
\be
m_1^2 + \frac{2\kappa_1}{g^2}\Omega^2 = m_2^2 + \frac{2\kappa_2}{g^2}\Omega^2
\ee

The 3-type solution we have found in Eqs.~\eqref{gaugeField}, \eqref{finalphi} for a uniform
electric field in SU(3) is very similar to the solution in SU(2). The main difference is that the
solution in SU(2) only required one scalar field, while in SU(3) we need at least two scalar fields.

\subsection{Electric string solution}
\label{Estringsoln}

In the previous subsection, we derived the uniform BW electric field solution for all equations of motion.
We now build on the uniform solution to construct electric string (flux tube) solutions 
by attaching a profile function $f(r)$ to the gauge field configuration \eqref{gaugeField} where 
$r$ is the cylindrical radial coordinate. 
Accordingly, we write
\be
W^{1}_{\mu} = -\frac{\epsilon}{g}\cos{\Omega t}f(r)\partial_{\mu}z, \ 
W^{2}_{\mu} = -\frac{\epsilon}{g}\sin{\Omega t}f(r)\partial_{\mu}z
\ee 
The non-vanishing field strength tensor components are
\ba
    W^{1}_{\mu\nu} &=& \frac{\epsilon}{g}\Omega \sin{\Omega t}f(r)(\partial_{\mu}t\partial_{\nu}z - \partial_{\nu}t\partial_{\mu}z) \nn \\
 && - \frac{\epsilon}{g}\cos{\Omega t}f'(r)(\partial_{\mu}r\partial_{\nu}z - \partial_{\nu}r\partial_{\mu}z) \\
    W^{2}_{\mu\nu} &=& -\frac{\epsilon}{g}\Omega \cos{\Omega t}f(r)(\partial_{\mu}t\partial_{\nu}z - \partial_{\nu}t\partial_{\mu}z) \nn \\
 &&
  - \frac{\epsilon}{g}\sin{\Omega t}f'(r)(\partial_{\mu}r\partial_{\nu}z - \partial_{\nu}r\partial_{\mu}z) 
\ea
and the non-vanishing gauge four-currents are,
\ba
\label{gaugecurrentstring1}
j^{1}_{\mu} &=& - \frac{\epsilon}{g}\cos{\Omega t}\Big[f''+\frac{f'}{r}+\Omega^2f\Big]\partial_{\mu}z \\
\label{gaugecurrentstring2}
j^{2}_{\mu} &=& - \frac{\epsilon}{g}\sin{\Omega t}\Big[f''+\frac{f'}{r}+\Omega^2f\Big]\partial_{\mu}z \\
\label{gaugecurrentstring3}
j^{3}_{\mu} &=& - \frac{\epsilon^2\Omega}{g}f^2\partial_{\mu}t
\ea

We include separate profile functions in each of the orthogonal scalar fields \eqref{finalphi},
\be
\label{stringphi}
\Phi_1 = \eta h_1(r)\begin{pmatrix}
    z_1 e^{i\omega t}\\ z_2 e^{-i\omega t}\\ 0
\end{pmatrix}, \
\Phi_2 = \eta h_2(r)\begin{pmatrix}
    w_1 e^{i\omega t}\\ w_2 e^{-i\omega t}\\ 0
\end{pmatrix}
\ee
Note that while $z_1, z_2, w_1, w_2$ satisfy the normalization and orthogonality constraints in \eqref{constraints}, $\eta$ and $\omega$ are arbitrary and will be fixed by the equations of motion.

The strategy now is to use the gauge field equations of motion to determine the gauge and scalar field 
profile functions. We will then use the scalar field equations of motion to obtain constraints on the 
parameters.

\begin{figure}
\includegraphics[width=0.45\textwidth,angle=0]{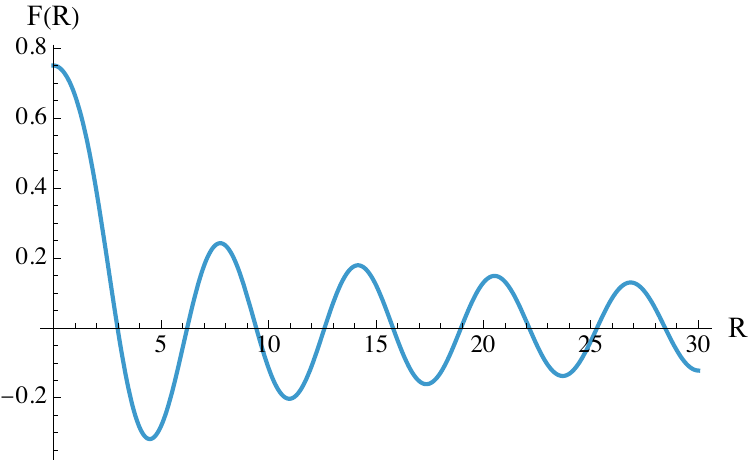}
 \caption{$F(R)$ vs $R$ for $F_0 = 0.75$}
\label{profile}
\end{figure}

We would like to solve the following set of equations
\be
\begin{aligned}
    j^{a}_{\mu} &= \sum_{i=1,2}\Big\{\frac{i g}{2} \big[\Phi_i^{\dagger}\lambda^{a}\partial_{\mu}\Phi_i - \text{h.c.}\big]\\ &+\frac{g^2}{3}W^{a}_{\mu}|\Phi_i|^2 + \frac{g^2}{2}d^{abc}W^{b}_{\mu}\big(\Phi_i^{\dagger}\lambda^c\Phi_i\big)\Big\}
\end{aligned}
\label{jeq}
\ee
where $j^{a}_{\mu}$ is defined in \eqref{gaugecurrentstring1}-\eqref{gaugecurrentstring3} and $\Phi_1$ and $\Phi_2$ are defined in \eqref{stringphi}. Substituting $\mu = t$ and $a=3$ gives
\be\label{gaugeeqn}
- g \omega \eta^2 \big(h_1^2 + h_2^2\big) = - \frac{\epsilon^2}{g}\Omega f^2,
\ee
and substituting $\mu = t$ and $a=8$ results in 
$h_1 = h_2 \equiv h$.
Substituting back in \eqref{gaugeeqn}, we obtain
\be\label{hdef}
2 g \omega \eta^2 h^2 = \frac{\epsilon^2}{g}\Omega f^2 \implies h(r) = \frac{\epsilon}{g\eta}\sqrt{\frac{\Omega}{2\omega}}f(r)
\ee
We must demand $\omega/\Omega > 0$ since $h$ is a real-valued function. 
For $\mu = z$ in \eqref{jeq}, we obtain
\be
f''+\frac{f'}{r} + \big(\Omega^2 - \eta^2g^2h^2\big)f = 0
\ee
Further on, substituting the definition of $h(r)$ from \eqref{hdef}, we can write
\be\label{fdef}
f''+\frac{f'}{r} + \Omega^2\Big(1 - \frac{\epsilon^2}{2\omega\Omega}f^2\Big)f = 0
\ee
This is the equation for the gauge profile function resulting from the gauge field equation of motion. We can plot this numerically by defining rescaled variables
\be
R = \Omega r,\quad F = \frac{\epsilon}{\sqrt{2\omega\Omega}}f
\ee
to write
\be
F''+\frac{F'}{R} + \big(1 - F^2\big)F = 0
\ee
Following \cite{Vachaspati:2023tpt}, we choose boundary conditions such that $F(0) = F_0 \neq 0$ and $F(\infty) \to 0$ along with $F'(0) = 0$. If $F_0 > 1$, then the $F^3$ term dominates and there are no solutions that fall off asymptotically as $R \to \infty$. Hence, we must restrict to $F_0 \leq 1$. A numerical plot of $F(R)$ for $F_0 = 0.75$ is shown in Fig.\ref{profile}.

\begin{figure}
\includegraphics[width=0.65\textwidth,angle=0]{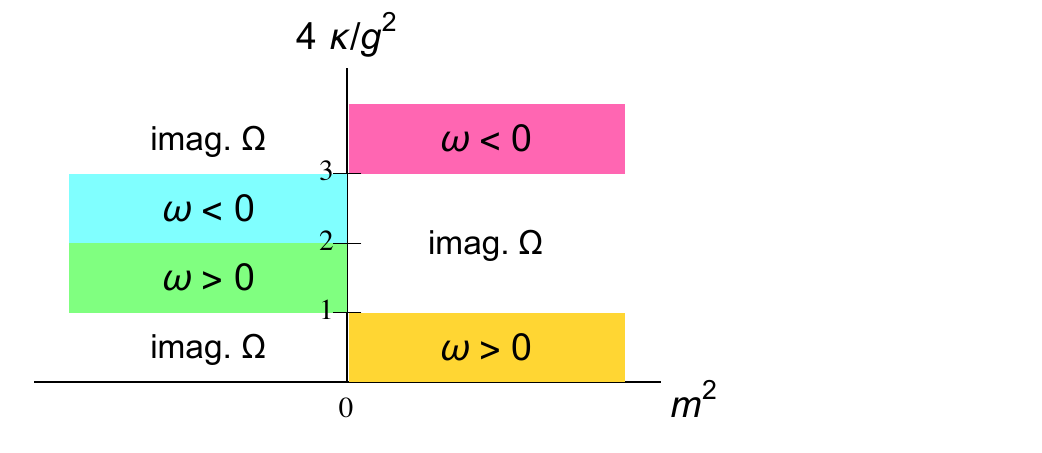}
 \caption{Allowed regions in parameter space spanned by different values of $m^2$ and $4\kappa/g^2$ are shaded. The solution is valid in regions with $\omega > 0$.}
\label{ParameterSpace}
\end{figure}

Now we consider the scalar field equation of motion, \eqref{Phieq}, to obtain
\be
h''+\frac{h'}{r} + \big(\omega^2-m^2\big)h-\frac{\epsilon^2}{4}f^2h-2\kappa\eta^2h^3 = 0
\ee
where we have set $m_1 = m_2 \equiv m$ and $\kappa_1 = \kappa_2 \equiv \kappa$. Substituting \eqref{hdef} and simplifying, we obtain
\be
f''+\frac{f'}{r} + \bigg[(\omega^2-m^2) - \bigg(\frac{\epsilon^2}{4} + \frac{\kappa\epsilon^2\Omega}{g^2\omega}\bigg)f^2\bigg]f = 0
\ee
Consistency with \eqref{fdef} implies
\ba
\label{consistency1}
\Omega^2 &=& \omega^2 - m^2\\
\label{consistency2}
\frac{\Omega}{2\omega} &=& \frac{1}{4} + \frac{\kappa\Omega}{g^2\omega}
\ea
We can solve \eqref{consistency2} to obtain
\be
\omega = 2 \bigg(1 - \frac{2\kappa}{g^2}\bigg)\Omega
\ee
As given below Eq.~\eqref{hdef}, we demand
$\omega /\Omega > 0$ which implies $0 \leq \kappa < g^2/2$. We can also solve \eqref{consistency1} and \eqref{consistency2} simultaneously to write
\ba
\Omega^2 &=& \frac{m^2}{4(1-2\kappa/g^2)^2 - 1} \\
\omega^2 &=& \frac{4(1-2\kappa/g^2)^2m^2}{4(1-2\kappa/g^2)^2 - 1}
\ea
Since $\Omega$ and $\omega$ are real, for $m^2 > 0$ we have 
$0\leq \kappa < g^2/4$, while for $m^2 < 0$ we have $g^2/4 < \kappa < g^2/2$ 
since $\kappa$ is bounded from above as discussed earlier. 
These allowed regions in parameter space are illustrated in Fig.~\ref{ParameterSpace}.

\subsection{Summary of electric string solution}
\label{summary}

Finally, we summarize the electric string solution:
\ba
\Phi_1 &=& \frac{\epsilon}{g}\sqrt{\frac{\Omega}{2\omega}}f(r)\begin{pmatrix}
    z_1 e^{i\omega t}\\ z_2 e^{-i\omega t}\\ 0
\end{pmatrix}, \\
\Phi_2 &=& \frac{\epsilon}{g}\sqrt{\frac{\Omega}{2\omega}}f(r)\begin{pmatrix}
    w_1 e^{i\omega t}\\ w_2 e^{-i\omega t}\\ 0
\end{pmatrix}\\
W^{1}_{\mu} &=& -\frac{\epsilon}{g}\cos{\Omega t}f(r)\partial_{\mu}z, \\ 
W^{2}_{\mu} &=& -\frac{\epsilon}{g}\sin{\Omega t}f(r)\partial_{\mu}z \\
W^{b}_{\mu} &=& 0 , \quad b = 3,\ldots,8
\ea
where
\be
\begin{aligned}
    |z_1|^2 + |z_2|^2 &= 1 \\ 
    |w_1|^2 + |w_2|^2 & = 1 \\
    z_1 w_1^{*} + z_2 w_2^{*} &= 0
\end{aligned}
\ee
along with
\ba
\Omega^2 &=& \frac{m^2}{4(1-2\kappa/g^2)^2 - 1} \\
\omega^2 &=& \frac{4(1-2\kappa/g^2)^2m^2}{4(1-2\kappa/g^2)^2 - 1}
\ea
The solution is valid for $0\leq \kappa < g^2/4$ for $m^2 > 0$ and $g^2/4 < \kappa < g^2/2$ for $m^2 < 0$. The profile function satisfies \eqref{fdef} along with $f(0) \leq \sqrt{2\omega\Omega}/\epsilon$ and $f'(0) = 0$.

\section{Schwinger pair production}
\label{unexciting}

In this section, we consider quantum fluctuations $Q^a_{\mu}$ about the background gauge 
fields (denoted $A^a_{\mu}$) that are responsible for the electric field. The motivation is to 
show that the electric field is stable to Schwinger particle production which, in temporal gauge, 
is equivalent to particle production in a time-dependent background~\cite{Vachaspati:2022ktr}.

Charged excitations in our SU(3) model include gauge particles and scalar particles,
and an electric field can Schwinger produce both kinds of particles. However, the mass
parameter of the scalar fields can be taken to be arbitrarily large, thus suppressing
Schwinger production of the scalar excitations. Hence, we will only consider
Schwinger pair production of gauge excitations, some of which can be massless.
Since the Schwinger process can be viewed as particle production in a time-dependent background, we only need to show that our electric field background is independent of time for the dynamics of the gauge particles. 

Consider perturbations about the gauge field background which give rise to electric fields in the $\lambda_3$ direction. 
The treatment is similar to that of \cite{Vachaspati:2022ktr}, however, we now need to consider perturbations in the $a=3,\ldots,8$ components as well. 
Consider
\be
W^a_{\mu} = A^a_{\mu} + Q^a_{\mu}
\ee
We can combine the fields to write them as follows
\ba
w^{\pm}_{\mu} &= a^{\pm}_{\mu} + e^{\mp i\Omega t}q^{\pm}_{\mu}\label{wdeflambda3}\\
y^{\pm}_{\mu} &= b^{\pm}_{\mu} + e^{\mp i\Omega t / 2}s^{\pm}_{\mu}\label{ydeflambda3}\\
z^{\pm}_{\mu} &= c^{\pm}_{\mu} + e^{\pm i\Omega t / 2}t^{\pm}_{\mu}\label{zdeflambda3}
\ea
where we have defined
\ba
w^{\pm}_{\mu} &= W^2_{\mu} \pm iW^1_{\mu} \\
y^{\pm}_{\mu} &= W^5_{\mu} \pm iW^4_{\mu} \\
z^{\pm}_{\mu} &= W^7_{\mu} \pm iW^6_{\mu}
\ea
and the background fields now take the following form
\ba
a^{\pm}_{\mu} &=& \mp\dfrac{i\epsilon}{g}e^{\mp i\Omega t}f(r)\partial_{\mu}z \\
b^{\pm}_{\mu} &=& 0, \quad c^{\pm}_{\mu} = 0\\
A^3_{\mu} &=& 0, \quad A^8_{\mu} = 0
\ea
Substituting these backgrounds in \eqref{wdeflambda3}-\eqref{zdeflambda3} gives
\ba
w^{\pm}_{\mu} &=& e^{\mp i\Omega t}\bigg[\mp\dfrac{i\epsilon}{g}f(r)\partial_{\mu}z + q^{\pm}_{\mu}\bigg]\label{wdef2lambda3}\\
y^{\pm}_{\mu} &=& e^{\mp i\Omega t / 2}s^{\pm}_{\mu}\label{ydef2lambda3}\\
z^{\pm}_{\mu} &=& e^{\pm i\Omega t / 2}t^{\pm}_{\mu}\label{zdef2lambda3}\\
W^3_{\mu} &=& Q^{3}_{\mu}, \quad W^8_{\mu} = Q^8_{\mu}\label{W3deflambda3}
\ea
Next, we can compute the field strength tensor using \eqref{fieldstrength},
\ba
w^{\pm}_{\mu\nu} &=&\ W^2_{\mu\nu} \pm i W^1_{\mu\nu} \nn \\
    &=&\ \partial_{\mu}w^{\pm}_{\nu} - \partial_{\nu}w^{\pm}_{\mu} \pm ig w^{\pm}_{\mu}W^3_{\nu} \mp ig W^3_{\mu}w^{\pm}_{\nu} \nn \\ 
    && \hskip 1 cm
    + \frac{g}{2}y^{\pm}_{\mu}z^{\mp}_{\nu} - \frac{g}{2}z^{\mp}_{\mu}y^{\pm}_{\nu}
\label{wfieldstrength}
\ea
Substituting \eqref{wdef2lambda3}-\eqref{W3deflambda3} in \eqref{wfieldstrength} gives
\be
\begin{aligned}
    w^{\pm}_{\mu\nu} =&\ e^{\mp i\Omega t}\Big[a^{\pm}_{\mu\nu} + \big(\mathcal{D}_{\mu}q^{\pm}_{\nu} - \mathcal{D}_{\nu}q^{\pm}_{\mu}\big) \\
    & \pm i g \big(q^{\pm}_{\mu}Q^3_{\nu} - q^{\pm}_{\nu}Q^3_{\mu}\big) + \frac{g}{2}\big(s^{\pm}_{\mu}t^{\mp}_{\nu} - s^{\pm}_{\nu}t^{\mp}_{\mu}\big)\Big]
\end{aligned}
\ee
with
\be\begin{aligned}
    a^{\pm}_{\mu\nu} = -\frac{\epsilon\Omega}{g}f(r)&\big(\partial_{\mu}t \partial_{\nu}z - \partial_{\nu}t\partial_{\mu}z\big) \\
    & \mp \frac{i\epsilon}{g}f'(r)\big(\partial_{\mu}r\partial_{\nu}z - \partial_{\nu}r\partial_{\mu}z\big)
\end{aligned}
\ee
\be
    \mathcal{D}_{\mu}q^{\pm}_{\nu} =\ \partial_{\mu}q^{\pm}_{\nu} \mp i\Omega \partial_{\mu}t q^{\pm}_{\nu} +\epsilon f(r)\partial_{\mu}zQ^3_{\nu}
\ee
Similarly, we have
\ba
    y^{\pm}_{\mu\nu} &=&\ W^5_{\mu\nu} \pm iW^4_{\mu\nu} \nn \\
    &=&\ \partial_{\mu}y^{\pm}_{\nu} - \partial_{\nu}y^{\pm}_{\mu} - \frac{g}{2}w^{\pm}_{\mu}z^{\pm}_{\nu} 
    + \frac{g}{2}z^{\pm}_{\mu}w^{\pm}_{\nu} \nn \\
    && \hskip 1 cm
    \pm ig\big(y^{\pm}_{\mu}x^{+}_{\nu} - x^{+}_{\mu}y^{\pm}_{\nu}\big)
\label{yfieldstrength}
\ea
where we have defined $x^{+}_{\mu} = \big(W^3_{\mu} + \sqrt{3}W^8_{\mu}\big)/2$.
Substituting \eqref{wdef2lambda3}-\eqref{W3deflambda3} in \eqref{yfieldstrength} gives
\be\begin{aligned}
    y^{\pm}_{\mu\nu} =&\ e^{\mp i\Omega t / 2}\Big[\big(\mathcal{D}_{\mu}s^{\pm}_{\nu} - \mathcal{D}_{\nu}s^{\pm}_{\mu}\big)\\
    & -\frac{g}{2}\big(q^{\pm}_{\mu}t^{\pm}_{\nu} - t^{\pm}_{\mu}q^{\pm}_{\nu}\big) \pm ig \big(s^{\pm}_{\mu}x^{+}_{\nu}-x^{+}_{\mu}s^{\pm}_{\nu}\big)
    \Big]
\end{aligned}\ee
where
\be
    \mathcal{D}_{\mu}s^{\pm}_{\nu} =\ \partial_{\mu}s^{\pm}_{\nu} \mp \frac{i\Omega}{2} \partial_{\mu}t s^{\pm}_{\nu} \pm \frac{i\epsilon}{2}f(r)\partial_{\mu}zt^{\pm}_{\nu}
\ee
Similarly,
\ba
    z^{\pm}_{\mu\nu} &=&\ W^7_{\mu\nu} \pm iW^6_{\mu\nu} \nn \\
    &=&\ \partial_{\mu}z^{\pm}_{\nu} - \partial_{\nu}z^{\pm}_{\mu} 
    + \frac{g}{2} w^{\mp}_{\mu}y^{\pm}_{\nu} - \frac{g}{2}y^{\pm}_{\mu}w^{\mp}_{\nu} \nn \\
    && \hskip 1 cm
    \pm ig x^{-}_{\mu}z^{\pm}_{\nu} \mp ig z^{\pm}_{\mu}x^{-}_{\nu}
\label{zfieldstrength}
\ea
where $x^{-}_{\mu} = \big(W^3_{\mu} - \sqrt{3}W^8_{\mu}\big)/2$. 
Once again, substituting \eqref{wdef2lambda3}-\eqref{W3deflambda3} in \eqref{zfieldstrength} yields
\be\begin{aligned}
    z^{\pm}_{\mu\nu} =&\ e^{\pm i\Omega t / 2}\Big[\big(\mathcal{D}_{\mu}t^{\pm}_{\nu} - \mathcal{D}_{\nu}t^{\pm}_{\mu}\big) \\
    & + \frac{g}{2}\big(q^{\mp}_{\mu}s^{\pm}_{\nu} - q^{\mp}_{\nu}s^{\pm}_{\mu}\big) \pm ig (x^{-}_{\mu}t^{\pm}_{\nu} - t^{\pm}_{\mu}x^{-}_{\nu})\Big]
\end{aligned}\ee
where we have defined
\be
\mathcal{D}_{\mu}t^{\pm}_{\nu} =\ \partial_{\mu}t^{\pm}_{\nu} \pm \frac{i\Omega}{2} \partial_{\mu}t t^{\pm}_{\nu} \pm \frac{i\epsilon}{2}f(r)\partial_{\mu}z s^{\pm}_{\nu}
\ee
Following the same procedure, we can write
\ba
    W^3_{\mu\nu} &=&\ \partial_{\mu}W^3_{\nu} - \partial_{\nu}W^3_{\mu} 
    - \frac{ig}{2}\big(w^{+}_{\mu}w^{-}_{\nu} - w^{-}_{\mu}w^{+}_{\nu}\big) \nn \\
    && \hskip -1 cm
    - \frac{ig}{4}\big(y^{+}_{\mu}y^{-}_{\nu} - y^{-}_{\mu}y^{+}_{\nu}\big) 
    + \frac{ig}{4} \big(z^{+}_{\mu}z^{-}_{\nu} - z^{-}_{\mu}z^{+}_{\nu}\big)
\label{W3fieldstrength}
\ea
\be\begin{aligned}\label{W8fieldstrength}
    W^8_{\mu\nu} =&\ \partial_{\mu}W^8_{\nu} - \partial_{\nu}W^8_{\mu} - \frac{i\sqrt{3}g}{4}\big(y^{+}_{\mu}y^{-}_{\nu} - y^{-}_{\mu}y^{+}_{\nu}\big) \\ & -\frac{i\sqrt{3}g}{4}\big(z^{+}_{\mu}z^{-}_{\nu} - z^{-}_{\mu}z^{+}_{\nu}\big)
\end{aligned}\ee
where substituting \eqref{wdef2lambda3}-\eqref{W3deflambda3} in \eqref{W3fieldstrength} and \eqref{W8fieldstrength} gives
\be\begin{aligned}\label{W3fieldstrengthfinal}
    W^3_{\mu\nu} =&\ \mathcal{D}_{\mu}Q^3_{\nu} - \mathcal{D}_{\nu}Q^3_{\mu} - \frac{ig}{2}\big(q^{+}_{\mu}q^{-}_{\nu} - q^{-}_{\mu}q^{+}_{\nu}\big) \\
    &- \frac{ig}{4} \big(s^{+}_{\mu}s^{-}_{\nu} - s^{-}_{\mu}s^{+}_{\nu}) + \frac{ig}{4}\big(t^{+}_{\mu}t^{-}_{\nu} - t^{-}_{\mu}t^{+}_{\nu}\big)
\end{aligned}\ee
\be\begin{aligned}\label{W8fieldstrengthfinal}
    W^8_{\mu\nu} =&\ \mathcal{D}_{\mu}Q^8_{\nu} - \mathcal{D}_{\nu}Q^8_{\mu} - \frac{i\sqrt{3}g}{4}\big(s^{+}_{\mu}s^{-}_{\nu} - s^{-}_{\mu}s^{+}_{\nu}\big) \\ & -\frac{i\sqrt{3}g}{4}\big(t^{+}_{\mu}t^{-}_{\nu} - t^{-}_{\mu}t^{+}_{\nu}\big)
\end{aligned}\ee
respectively, where we have defined
\be
\mathcal{D}_{\mu}Q^3_{\nu} = \partial_{\mu}Q^{3}_{\nu} - \epsilon f(r)\partial_{\mu}z q^2_{\nu}
\ee
along with $\mathcal{D}_{\mu}Q^8_{\nu} = \partial_{\mu}Q^8_{\nu}$.

The gauge field Lagrangian density can then be written as
\be\begin{aligned}\label{lagrangian}
    \mathcal{L}_{g} =\ -\frac{1}{4}&W^a_{\mu\nu}W^{\mu\nu a} \\
    =-\frac{1}{4}\Big[&w^{+}_{\mu\nu}w^{\mu\nu -} + y^{+}_{\mu\nu}y^{\mu\nu -} + z^{+}_{\mu\nu}z^{\mu\nu -} \\
    &+ W^{3}_{\mu\nu}W^{\mu\nu 3} + W^{8}_{\mu\nu}W^{\mu\nu 8}\Big]
\end{aligned}
\ee
Substituting all the above field strength tensors in the gauge field Lagrangian density \eqref{lagrangian}, we can check that the explicit time dependence completely cancels out. 

This means that the dynamics of the gauge perturbations denoted by $q_\mu^\pm$, $s_\mu^\pm$, $t_\mu^\pm$, $Q_\mu^3$, and $Q_\mu^8$ occur in a time-independent background, and they will not be excited. Equivalently, their quanta will not be produced.

\section{Conclusions}
\label{conclusions}

We have constructed 3-type electric string solutions in SU(3) non-Abelian gauge theory, as
summarized in Sec.~\ref{summary}. The construction follows the procedure in 
Ref.~\cite{Vachaspati:2022gco,Vachaspati:2023tpt} for SU(2) theory but requires 
two scalar fields in the fundamental representation instead of just one. The construction of 8-type electric strings remains an open question.

Such electric solutions may potentially be relevant to non-Abelian gauge theories such as QCD, though the fundamental scalar fields would have to arise as effective degrees of freedom, perhaps due to backreaction of quantum excitations. 
A novelty of the 3-type electric string solutions, and also of the uniform 3-type electric field, is that they are stable to the quantum production of gauge bosons due to the Schwinger process.

There are a number of follow-up questions that we plan to address in the future.
Are the 3-type solutions stable as in the SU(2) case~\cite{Pereira:2024sic}? Can such
solutions be studied using lattice methods? Can fundamental scalar fields arise
as emergent degrees of freedom in pure non-Abelian gauge theory?

\acknowledgements
This work was supported by the U.S. Department of Energy, Office of High Energy Physics, under Award No.~DE-SC0019470.

\appendix
\section{No solution with scalar field in the adjoint representation}
\label{appC}

In this Appendix, we show that there exists no solution with the scalar fields in the adjoint representation. Instead of providing a general proof for this, we will choose a scalar field that satisfies certain consistency conditions and then show that the corresponding scalar field four-current does not satisfy the gauge field equations.

Consider the homogeneous electric field in the $\lambda_3$ direction. The non-vanishing $SU(3)$ gauge field components are (see \eqref{gaugeField})
\be
W^{1}_{\mu} = -\frac{\epsilon}{g}\cos{\Omega t}\partial_{\mu}z, \ \ 
W^{2}_{\mu} = -\frac{\epsilon}{g}\sin{\Omega t}\partial_{\mu}z
\ee 
and the corresponding non-vanishing four-current components are (see \eqref{gaugecurrent31}-\eqref{gaugecurrent33})
\ba
j^{1}_{\mu} &=& - \frac{\epsilon\Omega^2}{g}\cos(\Omega t)\partial_{\mu}z
\label{adjointgaugecurrent1}
\\
j^{2}_{\mu} &=& - \frac{\epsilon\Omega^2}{g}\sin(\Omega t)\partial_{\mu}z 
\label{adjointgaugecurrent2}
\\
j^{3}_{\mu} &=& - \frac{\epsilon^2\Omega}{g}\partial_{\mu}t
\label{adjointgaugecurrent3}
\ea

The four-current for the adjoint scalar $\phi^a$ is
\be\label{adjointscalarcurrent}
J^a_{\mu} = g f^{abc}\phi^b D_{\mu}\phi^c
\ee
where the covariant derivative on $\phi^a$ is now defined as
\be
D_{\mu}\phi^a = \partial_{\mu}\phi^a + g f^{abc} W^b_{\mu}\phi^c
\ee
Since $f^{abc}$ is totally anti-symmetric, one straightforward constraint on $\phi^a$ is 
\be\label{adjointconstraint}
\phi^aJ^a_{\mu} = 0
\ee
and repeated indices $a,b,c = 1,\ldots,8$ are summed over.
Setting $\mu = t$ gives $\phi^{3} = 0$ while setting $\mu = z$ gives
\be
\cos{(\Omega t)}\phi^{1} + \sin{(\Omega t)}\phi^{2} = 0
\ee
Hence, we can choose the non-vanishing components of $\phi^a$ as
\be
\phi^1 = -\eta \sin{(\Omega t)}, \ \ \phi^2 = \eta \cos{(\Omega t)}
\ee
Using \eqref{adjointscalarcurrent}, the non-vanishing components of the scalar field four-current are
\ba
J^{1}_{\mu} &=& - \epsilon g \eta^2\cos(\Omega t)\partial_{\mu}z
\\
J^{2}_{\mu} &=& - \epsilon g \eta^2\sin(\Omega t)\partial_{\mu}z 
\\
J^{3}_{\mu} &=& g \eta^2\Omega\partial_{\mu}t
\ea
Comparing with \eqref{adjointgaugecurrent1}-\eqref{adjointgaugecurrent3} gives
\be
\eta = \frac{\Omega}{g}
\ee
and 
\be
\epsilon^2 = -\Omega^2
\ee
Since both $\epsilon$ and $\Omega$ are real, there is no consistent non-trivial solution with scalar fields in the adjoint representation.

\bibstyle{aps}
\bibliography{paper}

\begin{thebibliography}{11}
\expandafter\ifx\csname natexlab\endcsname\relax\def\natexlab#1{#1}\fi
\expandafter\ifx\csname bibnamefont\endcsname\relax
  \def\bibnamefont#1{#1}\fi
\expandafter\ifx\csname bibfnamefont\endcsname\relax
  \def\bibfnamefont#1{#1}\fi
\expandafter\ifx\csname citenamefont\endcsname\relax
  \def\citenamefont#1{#1}\fi
\expandafter\ifx\csname url\endcsname\relax
  \def\url#1{\texttt{#1}}\fi
\expandafter\ifx\csname urlprefix\endcsname\relax\def\urlprefix{URL }\fi
\providecommand{\bibinfo}[2]{#2}
\providecommand{\eprint}[2][]{\url{#2}}

\bibitem[{\citenamefont{Brown and Weisberger}(1979)}]{BROWN1979285}
\bibinfo{author}{\bibfnamefont{L.~S.} \bibnamefont{Brown}} \bibnamefont{and} \bibinfo{author}{\bibfnamefont{W.~L.} \bibnamefont{Weisberger}}, \bibinfo{journal}{Nuclear Physics B} \textbf{\bibinfo{volume}{157}}, \bibinfo{pages}{285} (\bibinfo{year}{1979}), ISSN \bibinfo{issn}{0550-3213}.

\bibitem[{\citenamefont{Casher et~al.}(1979)\citenamefont{Casher, Neuberger, and Nussinov}}]{Casher1979}
\bibinfo{author}{\bibfnamefont{A.}~\bibnamefont{Casher}}, \bibinfo{author}{\bibfnamefont{H.}~\bibnamefont{Neuberger}}, \bibnamefont{and} \bibinfo{author}{\bibfnamefont{S.}~\bibnamefont{Nussinov}}, \bibinfo{journal}{Phys. Rev. D} \textbf{\bibinfo{volume}{20}}, \bibinfo{pages}{179} (\bibinfo{year}{1979}).

\bibitem[{\citenamefont{Nayak and van Nieuwenhuizen}(2005)}]{Nayak2005}
\bibinfo{author}{\bibfnamefont{G.~C.} \bibnamefont{Nayak}} \bibnamefont{and} \bibinfo{author}{\bibfnamefont{P.}~\bibnamefont{van Nieuwenhuizen}}, \bibinfo{journal}{Phys. Rev. D} \textbf{\bibinfo{volume}{71}}, \bibinfo{pages}{125001} (\bibinfo{year}{2005}), \eprint{hep-ph/0504070}.

\bibitem[{\citenamefont{Cardona and Vachaspati}(2021)}]{Cardona:2021ovn}
\bibinfo{author}{\bibfnamefont{C.}~\bibnamefont{Cardona}} \bibnamefont{and} \bibinfo{author}{\bibfnamefont{T.}~\bibnamefont{Vachaspati}}, \bibinfo{journal}{Phys. Rev. D} \textbf{\bibinfo{volume}{104}}, \bibinfo{pages}{045009} (\bibinfo{year}{2021}), \eprint{2105.08782}.

\bibitem[{\citenamefont{Pereira and Vachaspati}(2025)}]{Pereira:2024sic}
\bibinfo{author}{\bibfnamefont{J.}~\bibnamefont{Pereira}} \bibnamefont{and} \bibinfo{author}{\bibfnamefont{T.}~\bibnamefont{Vachaspati}}, \bibinfo{journal}{Phys. Rev. D} \textbf{\bibinfo{volume}{111}}, \bibinfo{pages}{056022} (\bibinfo{year}{2025}), \eprint{2412.05458}.

\bibitem[{\citenamefont{Pereira and Vachaspati}(2022)}]{Pereira:2022lbl}
\bibinfo{author}{\bibfnamefont{J.}~\bibnamefont{Pereira}} \bibnamefont{and} \bibinfo{author}{\bibfnamefont{T.}~\bibnamefont{Vachaspati}}, \bibinfo{journal}{Phys. Rev. D} \textbf{\bibinfo{volume}{106}}, \bibinfo{pages}{096019} (\bibinfo{year}{2022}), \eprint{2207.05102}.

\bibitem[{\citenamefont{Vachaspati}(2022)}]{Vachaspati:2022ktr}
\bibinfo{author}{\bibfnamefont{T.}~\bibnamefont{Vachaspati}}, \bibinfo{journal}{Phys. Rev. D} \textbf{\bibinfo{volume}{105}}, \bibinfo{pages}{105011} (\bibinfo{year}{2022}), \eprint{2204.01902}.

\bibitem[{\citenamefont{Vachaspati}(2023{\natexlab{a}})}]{Vachaspati:2022gco}
\bibinfo{author}{\bibfnamefont{T.}~\bibnamefont{Vachaspati}}, \bibinfo{journal}{Phys. Rev. D} \textbf{\bibinfo{volume}{107}}, \bibinfo{pages}{L031903} (\bibinfo{year}{2023}{\natexlab{a}}), \eprint{2212.00808}.

\bibitem[{\citenamefont{Vachaspati}(2023{\natexlab{b}})}]{Vachaspati:2023tpt}
\bibinfo{author}{\bibfnamefont{T.}~\bibnamefont{Vachaspati}}, \bibinfo{journal}{Phys. Rev. D} \textbf{\bibinfo{volume}{107}}, \bibinfo{pages}{096015} (\bibinfo{year}{2023}{\natexlab{b}}), \eprint{2303.03459}.

\bibitem[{\citenamefont{Haber}(2021)}]{Haber}
\bibinfo{author}{\bibfnamefont{H.~E.} \bibnamefont{Haber}}, \bibinfo{journal}{SciPost Phys. Lect. Notes} p.~\bibinfo{pages}{21} (\bibinfo{year}{2021}).

\bibitem[{\citenamefont{Rubakov}(2002)}]{Rubakov:2002fi}
\bibinfo{author}{\bibfnamefont{V.~A.} \bibnamefont{Rubakov}}, \emph{\bibinfo{title}{{Classical theory of gauge fields}}} (\bibinfo{publisher}{Princeton University Press}, \bibinfo{address}{Princeton, New Jersey}, \bibinfo{year}{2002}), ISBN \bibinfo{isbn}{978-0-691-05927-3, 978-0-691-05927-3}.

\end{thebibliography}

\end{document}